\documentclass[twocolumn,prc,superscriptaddress,showpacs,preprintnumbers,amsmath,amssymb]{revtex4}
\usepackage{graphicx}
\usepackage{dcolumn}
\usepackage{bm}     

\usepackage[T1]{fontenc}        
\usepackage[latin1]{inputenc}   
\usepackage{times}              
\usepackage{amssymb}            %
\usepackage{amsmath}            
\usepackage{bbm}                
\usepackage{stmaryrd}           
\usepackage{xspace}             

\newcommand{\bc} { \begin{center}}
\newcommand{\ec} { \end{center}}
\newcommand{\br} { \begin{flushright}}
\newcommand{\er} { \end{flushright}}
\newcommand{\bl} { \begin{flushleft}}
\newcommand{\el} { \end{flushleft}}
\newcommand{\ba} { \begin{array}}
\newcommand{\ea} { \end{array}}
\newcommand{\be} { \begin{equation}}
\newcommand{\ee} { \end{equation}}
\newcommand{\bi} { \begin{itemize}}
\newcommand{\ei} { \end{itemize}}
\newcommand{\bea}{ \begin{eqnarray}}
\newcommand{\eea}{ \end{eqnarray}}

\newcommand{\Fig} {Fig.\ }

\newcommand{\Tab} {Tab.\ }

\newcommand{\Sec} {Section\xspace}

\newcommand{\textlatin}[1]{\textit{#1}\xspace}
\newcommand{\ie}    {\textlatin{i.e.}}
\newcommand{\eg}    {\textlatin{e.g.}}
\newcommand{\al}    {\textlatin{et al.}}

\newcommand{\p}[1]  {\phantom{#1}}
\begin{document}
\preprint{}
\title{Evidence for a breakdown of the Isobaric Multiplet Mass Equation:\\
A study of the $A=35$, $T=3/2$ isospin quartet}

\author{C.~Yazidjian}
 \altaffiliation[Corresponding address: ]
 {CERN, Physics Department, 1211 Geneva 23, Switzerland\\
 Electronic address: Chabouh.Yazidjian@CERN.ch\\
 This publication comprises part of the PhD thesis of C. Yazidjian}
\affiliation{GSI, Planckstra{\ss}e 1, 64291 Darmstadt, Germany}

\author{G.~Audi}
\affiliation{CSNSM-IN2P3-CNRS, 91405 Orsay-Campus, France}

\author{D.~Beck}%
\affiliation{GSI, Planckstra{\ss}e 1, 64291 Darmstadt, Germany}

\author{K.~Blaum}
\affiliation{GSI, Planckstra{\ss}e 1, 64291 Darmstadt, Germany}
\affiliation{Institut f{\"u}r Physik, Johannes Gutenberg-Universit{\"a}t, 55128 Mainz, Germany}

\author{S.~George}
\affiliation{GSI, Planckstra{\ss}e 1, 64291 Darmstadt, Germany}
\affiliation{Institut f{\"u}r Physik, Johannes Gutenberg-Universit{\"a}t, 55128 Mainz, Germany}

\author{C.~Gu\'enaut}
\altaffiliation[Present address: ]{NSCL, Michigan State University, East Lansing, MI 48824-1321, USA}
\affiliation{CSNSM-IN2P3-CNRS, 91405 Orsay-Campus, France}

\author{F.~Herfurth}
\affiliation{GSI, Planckstra{\ss}e 1, 64291 Darmstadt, Germany}

\author{A.~Herlert}
\altaffiliation[Present address: ]{CERN, Physics Department, 1211 Geneva 23, Switzerland}
\affiliation{Institut f{\"u}r Physik, Ernst-Moritz-Arndt-Universit{\"a}t, 17487 Greifswald, Germany}

\author{A.~Kellerbauer}
\altaffiliation[Present address: ]{Max Planck Institute for Nuclear Physics, P.O. Box 103980, 69029 Heidelberg, Germany}
\affiliation{CERN, Physics Department, 1211 Geneva 23, Switzerland}

\author{H.-J.~Kluge}
\affiliation{GSI, Planckstra{\ss}e 1, 64291 Darmstadt, Germany}
\affiliation{Ruprecht-Karls-Universit\"at, Institut f\"ur Physik, 69120 Heidelberg, Germany}

\author{D.~Lunney}
\affiliation{CSNSM-IN2P3-CNRS, 91405 Orsay-Campus, France}

\author{L.~Schweikhard}
\affiliation{Institut f{\"u}r Physik, Ernst-Moritz-Arndt-Universit{\"a}t, 17487 Greifswald, Germany}

\date{\today}

\begin{abstract}
Mass measurements on radionuclides along the potassium isotope chain have been performed with the
ISOLTRAP Penning trap mass spectrometer. For $^{35}$K ($T_{1/2}$ = 178 ms) to $^{46}$K ($T_{1/2}$
= 105 s) relative mass uncertainties of 2$\times 10^{-8}$ and better have been achieved. The
accurate mass determination of $^{35}$K ($\delta m$ = 0.54 keV) has been exploited to test the
Isobaric Multiplet Mass Equation (IMME) for the $A=35$, $T=3/2$ isospin quartet. The experimental
results indicate a deviation from the generally adopted quadratic form.%
\end{abstract}


\maketitle

\section{Introduction}

The application of the isospin formalism in nuclear physics stems from the assumption that the
strong interaction is almost charge-independent. In addition to the approximation that the neutron
and the proton have the same mass, the isospin formalism describes the neutron and the proton as
identical particles with isospin $T=1/2$ with the projections $T_\mathrm{z}(n)$ = $+1/2$ and
$T_\mathrm{z}(p)$ = $-1/2$, respectively \cite{Heis32,Wign37}. Isobaric nuclei with the same
isospin $T$ belong to a $2T+1$ multiplet with the projections $T_\mathrm{z} = (N-Z)/2$, where $N$
is the number of neutrons and $Z$ the number of protons in the nucleus. The corresponding states of
these isobars with the same $J^\pi$ and isospin $T$ are called Isobaric Analog States (IAS). The
IAS have nearly the same wave function, mainly perturbed by the charge difference in the nuclei.
Assuming only a two-body Coulomb force for the perturbation, the energy shift due to the
charge-state difference can be calculated. In first-order perturbation theory, this approximation
leads to a quadratic equation \cite{Wign58,Wein59}:
\begin{equation} \label{Eq:IMME}
D(A,T,T_\mathrm{z})\ =\ a(A,T)\ +\ b(A,T)T_\mathrm{z}\ +\ c(A,T)T_\mathrm{z}^2,
\end{equation}
that gives the mass excess $D$ of a member of a multiplet as a function of its atomic mass
number $A$, its isospin $T$ and isospin projection $T_\mathrm{z}$. This relation is known as the
Isobaric Multiplet Mass Equation (IMME).

The IMME can be used as a local mass model to predict unknown masses where some members of a
multiplet are known. Short-range predictions can provide accurate masses with uncertainties as low
as a few keV in favorable cases \cite{Lunn03}. This can be very helpful for applications such as
nuclear astrophysics, in particular, modeling the rapid proton-capture (rp) process where such
local models are in fact preferred to global models \cite{Scha06}. 

\begin{figure}[!h]
\begin{center}
\resizebox{0.48\textwidth}{!}{\includegraphics{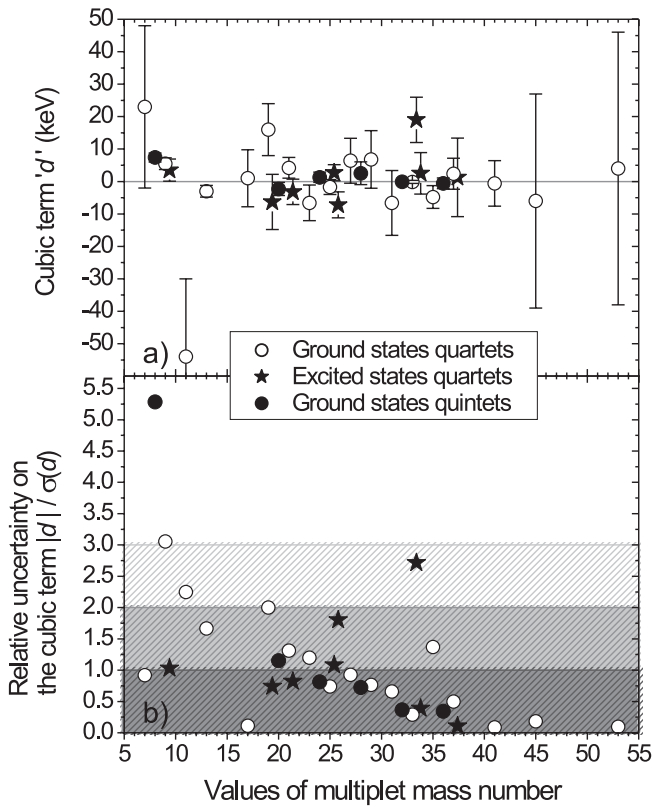} }
\end{center}
\caption{Compilation for the cubic parameter $d$ of the IMME \cite{Brit98,Herf01,Pyle02,Blau03a}.
Top: Values of $d$ as a function of the mass value $A$. Bottom: Relative uncertainty of $d$ where
the shaded areas indicate the 1, 2 and 3 $\sigma$ agreement.
} %
\label{fig:IMME}
\end{figure}

The IMME has been thoroughly studied in the late 70s \cite{Bene79}. Since then, many additional
measurements and tests have been performed and reported (see \eg \cite{Herf01,Pyle02,Blau03a}).
The IMME was found to work very well for
most cases. However, from the latest data compilation \cite{Brit98} and recent results of
experiments, some cases show a deviation from the quadratic form of the IMME
and need additional higher order terms \cite{Bert70}. Tests require
systems with at least four nuclides in the multiplet, \ie, with an isospin $T\geq 3/2$. Up to now,
only the $A=9$, $T=3/2$ quartet as well as the $A=8$, $T=2$ quintet system are known to deviate
significantly, \ie, by more than three standard deviations, from the quadratic form of the IMME
(see \Fig \ref{fig:IMME}). For those multiplets higher order terms have to be added, either
$dT_\mathrm{z}^3$, $eT_\mathrm{z}^4$, or both. The present paper reports on the improvement of the
$^{35}$K mass, which allows a further test of the quadratic form of the IMME for the $A=35$,
$T=3/2$ isospin quartet.

\section{The ISOLTRAP experiment}
\label{isoltrap}

The tandem Penning trap mass spectrometer ISOLTRAP \cite{Boll96}, installed at the on-line isotope
separator ISOLDE \cite{Krug01} at CERN (Geneva), is an experiment dedicated to high-precision
mass measurements on short-lived radionuclides \cite{Schw05,Blau06}. The mass measurement is based
on the determination of the cyclotron frequency $\nu_{\mathrm{c}}$ of a stored ion, probed by use of
radiofrequency (rf) fields in a Penning trap. With a charge-to-mass ratio $q/m$ and a magnetic
field $B$ the cyclotron frequency is given by:
\begin{equation}
\label{Eq:nu_c} %
\nu_\mathrm{c}= \frac{1}{2\pi}\frac{q}{m} B .
\end{equation}
With this technique, a relative mass uncertainty of the order of $\delta m/m = 10^{-8}$ is
routinely reached with ISOLTRAP~\cite{Kell03}. 

\begin{figure}[!ht]
\begin{center}
\resizebox{0.48\textwidth}{!}{\includegraphics{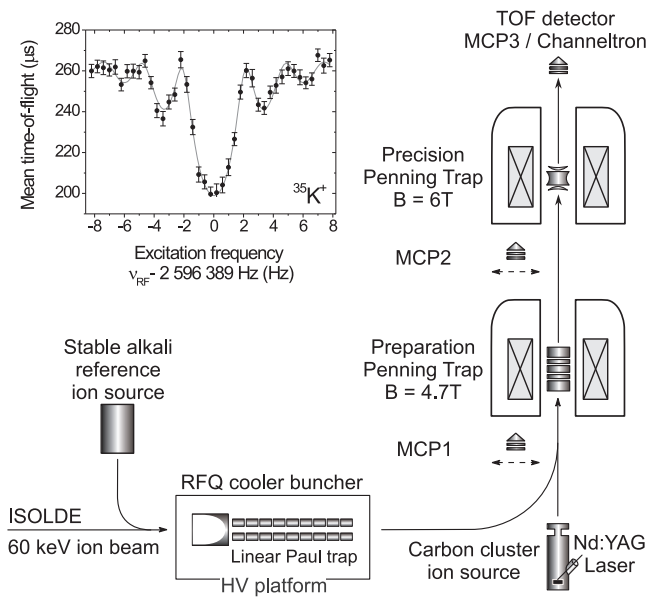}}
\end{center}
\caption{Sketch of the experimental setup of the ISOLTRAP mass spectrometer. Micro Channel Plate
(MCP) detectors are used to monitor the ion-beam transfer (MCP1-2). A newly implemented Channeltron
detector \cite{Yazi06}, which can be used in place of MCP3, records the time-of-flight (TOF)
resonance curve for the cyclotron frequency determination. The inset shows a resonance curve for
$^{35}$K$^+$ (excitation duration 400\,ms) together with the fit of the
theoretically expected line shape to the data points \cite{Koen95}.} 
\label{fig:ISOLTRAP_K35}
\end{figure}

The setup is composed of mainly three parts (see \Fig \ref{fig:ISOLTRAP_K35}): First, a linear
radiofrequency quadrupole (RFQ) cooler and buncher has the task to stop, accumulate, cool, and bunch the
60-keV ISOLDE beam for a subsequent transfer into the preparation trap \cite{Herf01b}. Second, the
cylindrical preparation Penning trap with helium buffer gas is used to remove isobaric contaminants
\cite{Sava91} with a resolving power $R = m/\Delta m$ up to 10$^5$ and to bunch the ions for an
efficient delivery to the second Penning trap. Finally, in the hyperbolical precision Penning trap
the cyclotron frequency $\nu_{\mathrm{c}}$ is determined for both the ion of interest and a
well-known reference ion by use of a quadrupolar rf-excitation \cite{Blau03}, for which the
frequency is varied around $\nu_{\mathrm{c}}$. The duration of the rf-excitation is chosen between
0.1 and 9\,s depending on the required resolution and the half-life of the ion of interest. As an
example of the time-of-fight technique \cite{Graf80} used at ISOLTRAP, the inset in \Fig
\ref{fig:ISOLTRAP_K35} shows a cyclotron resonance for $^{35}$K$^+$.

\begin{table*}[!ht]
\center%
\caption{Frequency ratios relative to the $^{39}$K$^+$ reference ion and mass-excess values of the investigated potassium isotopes.
The mass excess $D_\mathrm{exp}$ is calculated by $D_\mathrm{exp}=M_{exp}-A$, where $A$ is the respective mass number
and $M_\mathrm{exp}$ the atomic mass as deduced from the experimentally determined frequency ratio:
$M_\mathrm{exp}=(M^{\mathrm{ref}}-m_\mathrm{e})\nu_{\mathrm{c}}^{\mathrm{ref}}/\nu_{\mathrm{c}}+m_\mathrm{e}$.}
\begin{ruledtabular}\begin{tabular}{ccccc c c}
Isotope &
{$T{_{1/2}}$\footnote{Values from \cite{AME03}.}} &
{Frequency ratio \footnote{Using $^{39}$K$^+$ as a reference.} 
$\nu_{\mathrm{c}}^{\mathrm{ref}}/\nu_{\mathrm{c}}$} &
{$D{_{\mathrm{exp}}}$\footnote{M$^{\mathrm{ref}}$($^{39}$K)\,=\,38\,963\,706.68\,(20)\,$\mathrm{\mu}$u
\cite{nndc}, $m_\mathrm{e}=548\,579.911\,0\,(12)$\,nu \cite{Waps03}, and 1u\,=\,931\,494.009\,(7)\,keV~\cite{Waps03}.} (keV)} &
{$D{_{\mathrm{lit}}}$\footnotemark[1] (keV)} &
{$\Delta =D{_{\mathrm{lit}}} - D{_{\mathrm{exp}}}$ } (keV)\\
\hline 
\noalign{\smallskip}
$^{35}$K & 178  (8)  ms\p{i} & $0.897\,962\,555\,1\,(140)    $ & $-11\ 172.73\ $  $(54)$ & $-11\ 169\p{.20}$  $(20)    $ & $\p{\ \,-9017284}  3.7\p{230984}$ \\
$^{36}$K & 342  (2)  ms\p{i} & $0.923\,455\,783\,2\,(97)\p{1}$ & $-17\ 416.83\ $  $(39)$ & $-17\ 426\p{.20}$  $(8)\p{0}$ & $\p{3017284}  -9.2\p{730984}$ \\
$^{37}$K & 1.22 (1)  s\p{im} & $0.948\,917\,614\,6\,(84)\p{1}$ & $-24\ 800.45\ $  $(35)$ & $-24\ 800.20$      $(9)\p{0}$ & $\p{\ \,-3917284}  0.3\p{720984}$ \\
$^{38}$K & 7.64 (2)  min     & $0.974\,472\,667\,5\,(112)    $ & $-28\ 800.69\ $  $(45)$ & $-28\ 800.7\p{2}$  $(4)\p{0}$ & $\p{\ \,-3917284}  0.0\p{723984}$ \\
$^{43}$K & 22.3 (1)  h\p{im} & $1.102\,584\,811\,7\,(113)    $ & $-36\ 575.19\ $  $(46)$ & $-36\ 593\p{.20}$  $(9)\p{0}$ & $\p{390284}  -17.9\p{723084}$ \\
$^{44}$K & 22.1 (2)  min     & $1.128\,271\,956\,6\,(115)    $ & $-35\ 781.29\ $  $(47)$ & $-35\ 810\p{.20}$  $(40)    $ & $\p{390174}  -28.8\p{723094}$ \\
$^{45}$K & 17.3 (6)  min     & $1.153\,914\,244\,3\,(144)    $ & $-36\ 615.43\ $  $(57)$ & $-36\ 608\p{.20}$  $(10)    $ & $\p{\ \,-3901284}  7.4\p{723098}$ \\
$^{46}$K & 105  (10) s\p{im} & $1.179\,612\,625\,8\,(201)    $ & $-35\ 413.71\ $  $(76)$ & $-35\ 418\p{.20}$  $(16)    $ & $\p{3901728}  -4.3\p{720984}$ \\
\end{tabular}
\end{ruledtabular}
\label{Tab:K}
\end{table*}

For the production of  radioactive potassium isotopes a titanium-foil target was used. It consists of a
stack of thin titanium foils (30 $\mu$m each) for a total thickness of 19\,g$\cdot$cm$^{-2}$ and a
quantity of 50\,g of titanium. By bombardment with 1.4-GeV protons from the CERN Proton-Synchrotron
Booster on the ISOLDE target, the radioactive potassium isotopes were produced. After diffusing
out of the heated target they were surface-ionized by a hot tungsten ionizer. The potassium ions
were then accelerated to 60 keV and delivered to the ISOLTRAP experiment via the ISOLDE
High-Resolution Separator (HRS). During three days, the isotopic chain of potassium has been explored
from $^{35}$K for the neutron-deficient side, up to $^{46}$K for the neutron-rich side. A total
number of 29 cyclotron resonances for radioactive potassium isotopes have been recorded, together
with 39 resonances of the stable reference ion $^{39}$K$^+$. No ion contamination was observed
during the beam time. In the following the ISOLTRAP mass values of the potassium isotopes are
discussed, especially the $^{35}$K mass and its influence on the quadratic form of the IMME.

\section{Results}
\label{sec:results}

\begin{figure}[b]
\resizebox{0.48\textwidth}{!}{\includegraphics{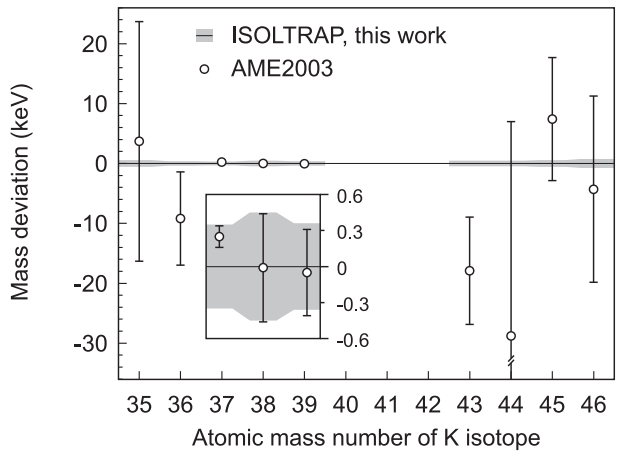}}%
\caption{Difference of mass-excess values ($D_\mathrm{lit}-D_\mathrm{exp}$) of the potassium isotopes taken
from this work ($D_\mathrm{exp}$, see Table\,\ref{Tab:K}) and the
literature \cite{AME03,nndc}. 
The experimental value of $^{39}$K results from a consistency check of the ISOLTRAP data (see text).
The shaded area represents the uncertainty of the ISOLTRAP values.
The differences of the mass-excess values of $^{37-39}$K are given as a cross-check.
For more details refer to the discussion section.} %
\label{Fig:K}
\end{figure}

Table \ref{Tab:K} summarizes the measured frequency ratios and the deduced mass-excess values of the
investigated isotopes. The mass determination for the potassium isotopes $^{35}$K ($T_{1/2}$ = 178
ms) up to $^{46}$K ($T_{1/2}$ =105 s) has been performed with an uncertainty below 2$\times
10^{-8}$, \ie, a reduction of the mass uncertainty by a factor of up to 40 for the
neutron-deficient side ($^{35}$K) and close to 80 for the neutron-rich potassium isotopes
($^{44}$K) as compared to the literature values \cite{AME03,nndc}. An overview is shown in
\Fig\ref{Fig:K}. The data for the well-known isotopes $^{37-39}$K are plotted as a cross-check for
the measurement process and the reliability of the ISOLTRAP results.

\begin{table*}[!ht]
\center%
\caption{Mass excess for nuclides of the $A=35$, $T=3/2$ quartet. The members of the $A=35$ quartet
with their respective half life and the associated isospin are given in the first three columns.
The corresponding mass excess of the ground state and the excitation energy and their uncertainty
for the IAS are given in columns 4 and 5. The last column indicates the total mass excess entering in
the IMME.}{}{}
\begin{ruledtabular}
\begin{tabular}{c c c c l   c }
\noalign{\smallskip} 
\multicolumn{1}{c}{Nucleus} & {$T{_{1/2}}$\footnote{Ground state, values from \cite{AME03}.}} &
$\ \ T{_{\mathrm{z}}}$ & 
\multicolumn{1}{c}{$D{_{\mathrm{exp}}^{\mathrm{gs }}}$\footnotemark[1] (keV)} & 
\multicolumn{1}{c}{$E{_{\mathrm{exp}}^{\mathrm{*}}}$\footnote{Values from \cite{Brit98}.} (keV)} & 
$D{_{\mathrm{exp}}^{\mathrm{tot}}}$ (keV) \\
\hline \noalign{\smallskip}
$^{35}$K\p{l}  & $178$ $(8)$  ms & $   - 3/2$ & $-11\ 172.73\ (0.54)$ \footnote{This work.} & \p{7777}---         $          $ & $-11\ 172.73\ (0.54)$ \\
$^{35}$Ar      & $1.775$ $(4)$ s &$   - 1/2$ & $-23\ 047.41\ (0.75)$ \p{\footnotesize\mbox{\texttt c}}   & $5\ 572.71 $ $(0.17)    $ & $-17\ 474.70\ (0.77)$ \\
$^{35}$Cl      & Stable &$\p{-}1/2$ & $-29\ 013.54\ (0.04)$ \p{\footnotesize\mbox{\texttt c}}             & $5\ 654\p{.17}$ $(2)$ & $-23\ 359.54\ (2.00)$ \\
$^{35}$S\p{l}  & $87.51$ $(12)$ d & $\p{-}3/2$ & $-28\ 846.36\ (0.10)$ \p{\footnotesize\mbox{\texttt c}}  & \p{7777}---         $          $ & $-28\ 846.36\ (0.10)$ \\
\noalign{\smallskip}
\end{tabular}
\end{ruledtabular}
\label{Tab:IMME1K35}
\end{table*}

\begin{table*}[!ht]
\center
\caption{Mass excess for nuclides of the $A=36$, $T=2$ quintet.The members of the $A=36$ quintet
with their respective half life and the associated isospin are given in the first three columns.
The corresponding mass excess of the ground state and the excitation energy and their uncertainty
for the IAS are given in columns 4 and 5. The last column indicates the total mass excess entering in
the IMME.}{}{}
\begin{ruledtabular}\begin{tabular}{c c c r l l }
\noalign{\smallskip} %
\multicolumn{1}{c}{Nucleus} & {$T{_{1/2}}$\footnote{Ground state, values from \cite{AME03}.}} &
$\ \ T{_{\mathrm{z}}}$ & %
\multicolumn{1}{c}{$D{_{\mathrm{exp}}^{\mathrm{gs }}}$ \footnotemark[1] (keV)} & %
\multicolumn{1}{c}{$E{_{\mathrm{exp}}^{\mathrm{*}}}$\footnote{Values from \cite{Endt90,Brit98,Endt98}.} (keV)} & %
$D{_{\mathrm{exp}}^{\mathrm{tot}}}$ (keV) \\
\hline \noalign{\smallskip}
$^{36}$Ca       & $102$ $(2)$ ms & $   - 2$ & $-\p{3}6\ 440\p{.86}\ (40)\p{.1}$  \p{\footnotemark[3]}\p{\footnotemark[3]} &  \p{7777} ---         $      $                      & $-\p{3}6\ 440\p{.91}\ (40)\p{.11}$ \\
$^{36}$K\p{l}   & $142$ $(2)$ ms & $   - 1$ & $-17\ 416.83\ (0.39)$   \footnote{This work.}\p{\footnotemark[6]}  & $\p{1} 4\ 282.2$ $(2.5)$ \footnote{Extracted from the ground state and the excited state value (see \footnotemark[5])\p{0}} & $-13\ 134.7\p{1}\ (2.4)$  \footnote{Value from \cite{Garc95} corrected for relativistic effect and for the $^{35}$Ar mass \cite{AME03}.}\p{0} \\ 
$^{36}$Ar       & Stable         & $\p{-}0$ & $-30\ 231.54\ (0.03)$   \footnote{See \Sec 7.4 in \cite{Waps03}.}\p{\footnotemark[3]}                 & $     10\ 851.6$ $(1.50)$                           & $-19\ 379.94\ (1.50)$ \\
$^{36}$Cl       & $301$ $(2)$ ky & $\p{-}1$ & $-29\ 521.86\ (0.07)$    \p{\footnotemark[3]}\p{\footnotemark[3]}           & $\p{1} 4\ 299.7$ $(0.08)$                           & $-25\ 222.16\ (0.11)$ \\
$^{36}$S\p{l}   & Stable         & $\p{-}2$ & $-30\ 664.07\ (0.19)$    \p{\footnotemark[3]}\p{\footnotemark[3]}           &  \p{7777} ---         $      $                      & $-30\ 664.07\ (0.19)$ \\
\noalign{\smallskip}%
\end{tabular}
\end{ruledtabular}
\label{Tab:IMME1K36}%
\end{table*}

The mass excess for $^{35}$K found in this work ($D{_{\mathrm{exp}}}=-11\,172.73\,(54)$\,keV)
agrees with the value given in the latest Atomic-Mass Evaluation (AME2003) \cite{AME03} but has a 40 times reduced uncertainty.
The consequences with respect to the quadratic form of the IMME are discussed in detail in the next
section.

Concerning $^{36}$K with a mass excess of $D_\mathrm{exp}=-17\,416.83\,(39)$\,keV, the only
contribution to the AME2003 arises from the $^{36}$Ar($p$,$n$)$^{36}$K reaction \cite{Goos71} and gives originally
$D=-17\,421\,(8)$\,keV. It has to be emphasized that the $(p,n)$ reaction energy
has been recalibrated afterwards \cite{Free76}. Other indirect mass determinations from
$^{36}$Ar($p$,$n$)$^{36}$K \cite{Jaff71} and $^{36}$Ar($^3$He,$t$)$^{36}$K \cite{Dzub70} can be
also found in the literature but were not used for the mass evaluation. Whereas the value given in
\cite{Jaff71} agrees within the uncertainty, a deviation of 2.4$\sigma$ is observed relative to the
value reported in \cite{Dzub70}. Finally, only a slight difference is observed compared to the
literature value from the recalibrated experiment of Goosman \al: $D_\mathrm{lit}=-17\,426.2\,(7.8)$\,keV
\cite{AME03} with 1.1$\sigma$. The mass of $^{36}$K has also an impact on the
IMME test for the $A=36$, $T=2$ quintet. The consequences for the quadratic form of the IMME are
presented together with the $A=35$, $T=3/2$ quartet in \Sec \ref{sec:IMMEdiscuss}.

The ISOLTRAP value for the mass excess of $^{37}$K ($D_\mathrm{exp}=-24\,800.45\,(35)$\,keV) has a four times larger
uncertainty than the literature value, since the latter is known with a precision better than the
current limit of our experiment $\sigma(m)/m=$8$\times$10$^{-9}$. However, the present result shows
an excellent agreement with the adopted mass-excess value, which gives strong confidence in the ISOLTRAP
results.

For $^{38}$K an isomeric state might have been present during the measurements.
The excitation energy of the isomeric state is well determined, $E^*=130.4\,(3)$\,keV
\cite{Endt90}, by measurements of the $^{38}$K$^m$(IT)$^{38}$K internal transition $\gamma$-rays,
after the production of $^{38}$K$^m$ with a $^{38}$Ar($p$,$n$)$^{38}$K$^m$ reaction. In specific
radioactive-beam preparation and together with laser ionization using the resonant laser
ionization method RILIS \cite{Kost02}, the ISOLTRAP experiment showed its ability to perform pure
isomeric mass determination in the case of $^{68}$Cu \cite{Blau04} and $^{70}$Cu \cite{Roos04}.
With an excitation time of $T_\mathrm{RF}=900$ ms, the resolving power was about 2$\cdot$10$^6$,
\ie, one order of magnitude higher than needed to resolve the two respective isomers. However, in
the present work only one of the isomeric states has been observed. For this reason, the resulting
mass could not be clearly assigned directly to any of the two isomeric states. The excited state
$^{38}$K$^m$ is shorter-lived (924 ms) than the ground state $^{38}$K (7.64 min). During the
cyclotron frequency determination procedure, radioactive nuclides may decay and produce
characteristic signals in the time-of-flight spectrum. The analysis of the cyclotron-resonance data
performed with an excitation time of 1.2 s did, however, not show any decay peaks. Moreover, the
obtained mass-excess value $-28\,800.69(45)$\,keV agrees with the literature value of the ground
state: $-28\,800.7\,(4)$\,keV \cite{AME03}. Therefore it can be concluded that the ground state
$^{38}$K and not the excited isomeric state was produced in the
ISOLDE target and directly measured with ISOLTRAP.

Even though $^{39}$K was used as the reference nuclide for all potassium measurements, all the
backward information flow from the investigated nuclides provided a contribution to a new mass
evaluation for $^{39}$K. In \cite{AME03}, the mass evaluation from all experimental data is done
by solving a system of linear equations. In the present work, the mass determination of the
well-known nuclides $^{37}$K and $^{38}$K, used for cross references and data consistency, slightly
changed the $^{39}$K mass-excess value ($D_\mathrm{exp+lit}=-33\,806.9\,(2)$\,keV, instead of
$D_\mathrm{lit}=-33\,807.0\,(2)$\,keV. This is of interest since $^{39}$K is used as a
reference mass in many experimental setups.

Previous work \cite{Lind54,Benc59} led to two different mass values  for $^{43}$K. The adopted
value in \cite{AME03}, which results from an average of those input data, has a large uncertainty
($D_\mathrm{lit}(^{43}$K$)=-36\,593\,(9)$\,keV). The value presented in this work shows a relative discrepancy of a
bit more than two standard deviations. The adopted literature value results from indirect mass
determinations. Sometimes such a deviation can be explained by missing $\gamma$-lines in the
recorded spectra. However, in the present case, no clear indication of missing levels has been
found.

Former investigations on the decay imply large uncertainties of the order of ten keV for $^{44}$K.
The comparison between the mass-excess value presented in this work and the original data shows a
good agreement with \cite{Ajze70} and a deviation of 1.3$\sigma$ from \cite{Levk70}. However, the
weighted average mass-excess value given in \cite{AME03} agrees with the ISOLTRAP data.

Concerning $^{45}$K and $^{46}$K, the uncertainties arise from the reactions
$^{46}$Ca($t$,$\alpha$)$^{45}$K \cite{Sant68} and $^{48}$Ca($d$,$\alpha$)$^{46}$K \cite{Mari65}.
Those articles are not well documented for a recalibration of the measurements (see \cite{Waps03}
and references therein). The mass-excess values presented in this work agree with the previous data
and improve the uncertainties by a factor of 20 to 80.

\section{Discussion}
\label{sec:IMMEdiscuss}

The two potassium isotopes $^{35}$K and $^{36}$K are involved in the $A=35$, $T=3/2$ isospin
quartet and the $A=36$, $T=2$ isospin quintet, respectively. In \Tab\ref{Tab:IMME1K35} and
\ref{Tab:IMME1K36}, the updated mass-excess values of the multiplets are summarized taking into
account the values presented in this work. Unfortunately, to fully evaluate the $A=36$ quintet, a
precise mass determination of $^{36}$Ca ($T_{1/2}= 102\,(2)$\,ms) is still missing
($D_\mathrm{lit}=-6440(40)$\,keV).

\begin{table}[!h]
\center%
\caption{The $A=35$, $T=3/2$ quartet and the coefficients for the quadratic and cubic form.}{}{}
\begin{tabular}{cccccc}
\hline  \hline \noalign{\smallskip}
$D(T_\mathrm{z})$ & $a$ (keV) & $b$ (keV) & $c$ (keV) & $d$ (keV) & $\chi_n^2$ \\
\hline \noalign{\smallskip}
Quadratic & $-$20 470.7(0.8)& $-$5 891.2(0.2) & 205.0(0.4)& ---     & 8.8\\
Cubic     & $-$20 468.1(0.2)& $-$5 884.0(2.4) & 203.8(0.6)& $-$3.2(1.1)& --- \\
\hline \hline \noalign{\smallskip}
\end{tabular}
\label{Tab:IMME2K35}%
\end{table}

\begin{table}[!h]
\center%
\caption{The $A=36$, $T=2$ quintet and the the coefficients for the quadratic and cubic form.}{}{}
\begin{tabular}{cccccc}
\hline  \hline \noalign{\smallskip}
$D(T_\mathrm{z})$ & $a$ (keV) & $b$ (keV) & $c$ (keV) & $d$ (keV) & $\chi_n^2$ \\
\hline \noalign{\smallskip}
Quadratic & $-$19 379.1(0.7)& $-$6 043.6(0.8) & 200.6(0.3)& ---     & 0.9\\
Cubic     & $-$19 380.3(1.5)& $-$6 043.3(1.1) & 202.1(1.8)& $-$0.7(0.8)& 1.1 \\
\hline \hline \noalign{\smallskip}
\end{tabular}
\label{Tab:IMME2K36}%
\end{table}

From the mass-excess values, the coefficients $a,\ b,\ c$ of the quadratic terms and the possible
coefficient $d$ of the cubic term of the IMME can be derived by use of a standard least mean square
fit. The results for the $A=35$, $T=3/2$ quartet, and the $A=36$, $T=2$ quintet, under the
assumption of quadratic and cubic forms of the IMME, are given in \Tab\ref{Tab:IMME2K35} and
\Tab\ref{Tab:IMME2K36}.

\begin{figure}[b]
\resizebox{.38\textwidth}{!}{\includegraphics{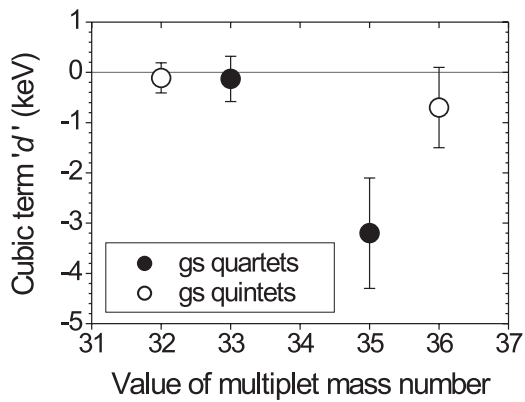}}%
\caption{Partially updated compilation of the coefficient $d$ of the cubic term for multiplets,
including this work. The four points shown correspond to the $A=33$ and $A=35$ quartets, and $A=32$
and $A=36$ quintets for which at least one member has been measured at ISOLTRAP
\cite{Herf01,Blau03a}. The two points
on the right hand side, \ie, the $A=35$ quartet and the $A=36$ quintet, arise from the present work.}%
\label{fig:IMME_3}
\end{figure}

The updated plot for the cubic term $d$ of the IMME is given in \Fig\ref{fig:IMME_3}. In the latest
compilation \cite{Brit98}, the $A=35$, $T=3/2$ isospin quartet was already reported to slightly
deviate, and the $A=36$-quintet followed the adopted quadratic form with a cubic term with the
coefficient $d=-0.6\,(1.6)$\,keV. In \cite{Brit98} the reduced $\chi^2$ for the quintet was close
to 3, and therefore indicated a possible non-consistent set of data.

However, it can be observed that for the $A=35$-quartet none of the coefficients of the quadratic
form agrees within one standard deviation with the corresponding coefficient for the cubic form.
This clearly indicates a strong discrepancy and is an argument in favor of using higher terms to
describe the IMME. For the $A=36$-quintet, which now shows no deviation within a standard
deviation, the uncertainties on the coefficients are larger, because of the lack of knowledge on
the $^{36}$Ca ground state. Unless a new high-precision mass measurement is performed, no final
conclusions about the quintet can be drawn. However, assuming the quadratic form of the IMME being
valid for the $A=36$, $T=2$ quintet, the mass value of $^{36}$Ca can be extrapolated from the other
members of the multiplet to $D_\mathrm{IMME}=-6490.3\,(6)$\,keV. This value slightly deviates
(1.3$\sigma$) from the previously adopted value \cite{AME03} but has a close to two orders of
magnitude smaller uncertainty.

In order to understand the reasons of the deviation observed for the $A=35$ quartet and to find out
the causes for higher-order terms in the IMME, it will be assumed that the quadratic form of the
IMME is correct and one (or more) of the ground state masses involved exhibit a systematic shift.
The same signature appears if one of the excited states is wrongly assigned.

\begin{table}[!h]
\center%
\caption{Mass prediction and residuals assuming a quadratic fit and using the coefficients as given
in \Tab\ref{Tab:IMME2K35}.}{}{}
\begin{ruledtabular}
\begin{tabular}{c cc c}
\multicolumn{4}{c}{$D(T_\mathrm{z}) = a + b T_\mathrm{z} + c T_\mathrm{z}^2 $}\\%
\hline \noalign{\smallskip}%
\multicolumn{1}{c}{Nucleus} & $D{_{\mathrm{cal}}^{\mathrm{tot}}}$&
$D{_{\mathrm{cal}}^{\mathrm{tot}}}-D{_{\mathrm{exp}}^{\mathrm{tot}}}$&  $E{_{\mathrm{cal}}^{\mathrm{*}}}$\\%
\multicolumn{1}{c}{       } & (keV) & (keV) &  (keV)\\%
\hline \noalign{\smallskip}
$^{35}$K\p{l}  & $-11172.9\ (1.2)$ & $    -0.2\ (1.3)$ &  --- \\%
$^{35}$Ar      & $-17473.9\ (0.8)$ & $\p{-}0.8\ (1.1)$ & $5573.5\ (1.1)$ \\%
$^{35}$Cl      & $-23365.0\ (0.8)$ & $    -5.5\ (2.1)$ & $5648.5\ (0.8)$ \\%
$^{35}$S\p{l}  & $-28846.4\ (1.2)$ & $\p{-}0.0\ (1.2)$ &  --- \\%
\end{tabular}
\end{ruledtabular}
\label{Tab:Predictabc}%
\end{table}

In \Tab\ref{Tab:Predictabc} the different mass excess predictions of the supposedly `unknown'
nuclides are shown, calculated with the fit parameters given in \Tab\ref{Tab:IMME2K35}. The
$^{35}$K, $^{35}$Ar, and $^{35}$S, values agree with the literature. However, compared to the value
given in \cite{Brit98,AME03} the $^{35}$Cl mass deviates by 2.6 standard deviations. Due to the
2\,keV uncertainty on the IAS, the $T_\mathrm{z}=1/2$ member  state in $^{35}$Cl has the least
significant contribution to the fit. The present status is identical to the IMME `breakdown'
reported in \cite{Herf01}, where the least significant member was $^{33}$Ar. The `revalidation' of
the IMME \cite{Pyle02} showed that the excited state of the $^{33}$Cl was erroneous. Therefore,
even if those direct mass extrapolation methods seem to indicate a deviation
of the excited $T_\mathrm{z}=1/2$ IAS for $^{35}$Cl, caution is advised.

In the following, a more detailed discussion is presented, where all ground state mass values for
the members of the $A=35$, $T=3/2$ quartet as well as the excitation energies for $^{35}$Cl and
$^{35}$Ar are included to find indications for any deviation of the IMME.

\subsection{Ground-state masses of the IMME $A=35$ quartet}
In the $A=35$, $T=3/2$ quartet four ground state masses are involved:
In this work the mass excess of $^{35}$K has been directly determined for the first time by a
Penning trap measurement technique. Only few cases showed a discrepancy to the Atomic-Mass Evaluation \cite{AME03}
that could not be resolved, as \eg, $^{36}$Ar (see \Sec 7.4 of \cite{Waps03} and references therein).
Moreover, the precision of the mass-excess value in this work is 40 times better than the adopted one
and agrees with it.

The mass excess of $^{35}$Ar results from an indirect mass measurement by means of the
$^{35}$Cl($p$,$n$)$^{35}$Ar reaction . Three input data values are taken \cite{Free75,Whit77,Azue78}. However, one of them
\cite{Azue78} deviates by 3.4 keV (close to two standard deviations) from the adopted value.
Nevertheless, this deviation alone is not sufficient to explain the discrepancy of the quadratic form of the IMME.

The mass of $^{35}$Cl has been determined by direct rf-measurements \cite{Smit71}, which
contribute about 79\% to the mass determination. Since the value steams from a direct
measurement, it is quite reliable and can be assumed to be correct.

The mass excess of $^{35}$S is mainly determined (95\%) by a $\beta$-endpoint measurement of the
$^{35}$S($\beta^-$)$^{35}$Cl
reaction, which was thoroughly studied for the presumed existence of a 17 keV-neutrino,
see \cite{Waps03} (\Sec 7.3 and references therein). Even though the data reported in \cite{Waps03}
are labeled as `well documented but not consistent with other well documented data', the
discrepancies observed are less than 0.4 keV
\cite{Altz85,Ohi85,Simp89,Chen92,Berm93,Mort93}. While those relatively small uncertainties and deviations
from the adopted value are  not sufficient to draw conclusions on the existence or absence of the 17
keV-neutrino, they are precise enough to presume the mass value of $^{35}$S is accurate, since no
systematic trends were found in the literature.

Thus, the careful study of the ground states did not show any deviation from the adopted values
\cite{AME03}, except maybe for $^{35}$Ar. In \cite{Waps03}, this nuclide is labeled as `secondary
data', \ie, where the mass is known from only one type of data, in the present case experimental
input from the $^{35}$Cl($p$,$n$)$^{35}$Ar reaction \cite{Free75,Whit77,Azue78}, and is not
cross-checked by a different connection.

\subsection{Excited states of the IMME $A=35$ quartet}
The values reported for the excited states are taken from \cite{Endt90,Brit98,Endt98,nndc} and
references therein. The adopted value for the IAS excited state of $^{35}$Ar does not show a strong
deviation from the experimental data. In addition, the different estimates for the excited state in
\Tab\ref{Tab:Predictabc} do not deviate by more than one standard deviation and are far from any
other known excited state in $^{35}$Ar. Therefore, it can be concluded that the excited state is
correctly assigned.

For the excited state of $^{35}$Cl, as summarized in \cite{Endt90}, the experimental data are not
precise enough and show as well some discrepancies (see \cite{Hube72a} and references therein). The
energy level scheme of $^{35}$Cl exhibits a `double' peak around 5.65 MeV, which has been
thoroughly investigated \cite{Wats67,Grau69,Hube72a,Fant73,Meye76}. Previous work on the IMME
showed that the energy of the excited state for $^{33}$Cl was wrongly calculated from the
center-of-mass to the laboratory frame \cite{Herf01}. From the raw data of the proton energy in the
laboratory frame \cite{Meye76}, the excitation energy has been recalculated taking into account
relativistic effects and compared to the values given in the above mentioned references. No major
deviation was found.

A detailed analysis of the excited state of $^{35}$Cl and a discussion of the separation and the
spin assignment of the two 5.65-MeV states can be found in \cite{Fant73}. The excitation energy
resulting from the mean values of the data gives the respective energies of 5646\,(2)\,keV and
5654\,(2)\,keV. The excited state at 5654\,(2)\,keV is the commonly adopted value for the $T=3/2$
IAS. However, the calculated excitation energy as given in \Tab\ref{Tab:Predictabc} 
corresponds to the excited state with energy 5646\,(2)\,keV. When using
this state rather than the adopted one, a cubic term $d=0.8\,(1.0)$\,keV is found in agreement with
zero. Therefore, it can be concluded that a misassignment of the IAS is possible.

Moreover, the $Q$-value found in \cite{Hube72a} shows a deviation of about 3 keV as compared to
\cite{AME03} for the $^{34}$S($p$,$\gamma$)$^{35}$Cl reaction. For the $A=33$ quartet \cite{Pyle02}
an unexpected shift of a few keV was revealed for the excited states of $^{33}$Cl. This shift was
sufficient to explain the observed `breakdown' of the IMME \cite{Herf01} and revalidated the
quadratic form of the IMME \cite{Pyle02}. Such a trend can also be the source of the deviation for
the $A=35$, $T=3/2$ quartet.

\subsection{Higher-order terms of the IMME $A=35$ quartet}

In the discussion above, it was assumed that the IMME has a pure quadratic form and an indication
for a possibly wrong mass and/or excitation energy has been found. Even though the adopted
quadratic form of the IMME is sufficient to describe the mass surface for a given multiplet,
experimental and theoretical studies pointed out the possibility of a deviation from the quadratic
form and the need for higher order terms in the IMME~\cite {Bert70}.

The excited-state assignment of $^{35}$Cl can be validated with a simulation based on a theoretical
model without isospin mixing \cite{Brow88}. The 3/2$^+$ excited state, corresponding to the IAS in
the $A=35$, $T=3/2$ quartet, shows a preferential branching ratio towards the 5/2$^+$ state lying
at 3 MeV. Compared to the decay scheme of bound states in $^{35}$Cl, where the 5654\,(2)\,keV and
5646\,(2)\,keV states decay towards the 5/2$^+$ and the 7/2$^-$ state, respectively, it can be
concluded that there is  no misassignment of the IAS in $^{35}$Cl. Calculations based on sd-shell
model calculations \cite{Orma89} with isospin dependent interaction show as well a deviation from
the quadratic form of the IMME with the same magnitude for the coefficient of
the cubic term but with the opposite sign,
\ie, $d=3.1$\,keV. The reasons for the sign difference are not clear yet but the IMME quadratic form
seems to be insufficient to describe the $A=35$, $T=3/2$ quartet from both the experimental and the
theoretical side.

The pure two-body Coulomb perturbation approximation to derive the IMME neglects the off-diagonal
part of the isovector and isotensor components of the Coulomb force. However, the latter might
introduce an isospin mixing which causes a shift in the levels of the different quartet members and
leads to a higher order polynomial form in $T_\mathrm{z}$.

It has been demonstrated in \cite{Henl69} that corrections to the quadratic form of the IMME can be
used. The correction with the coefficient $d$ of the cubic term is expected to be proportional to $Z\alpha c$, where
$Z$ is the proton number, $\alpha$ the fine-structure constant, and $c$\/ the coefficient of the
quadratic form of the IMME. However, the calculated $d$ values are found to be smaller than
$Z\alpha c$. This can be explained by the fact that the second-order corrections are `absorbed' in
the $a,b,c$ coefficients. Isospin violation of the nuclear interaction inducing a small
isospin-breaking component can also lead to higher order terms in the IMME. If the bare nuclear
interaction has a three-body component, and if it is isospin violating, it would automatically
 lead to a cubic $T_\mathrm{z}^3$ term. In the vicinity of the $A=35$, $T=3/2$ quartet
members unexpected isospin-breaking and -mixing effects have been recently observed for the 7/2$^-$
and 13/2$^-$ states between the $^{35}$Cl and $^{35}$Ar mirror nuclides \cite{Ekam04}. If the
isospin $T$ is a good quantum number, the $E1$ transitions are identical in mirror nuclei, which is
not the case. The reason is an isospin mixing of the $\left| {7/2^-} \right\rangle$ and $\left|
{5/2^+} \right\rangle$ levels.

\section{Conclusion} \label{conclu}
The thorough study of the $A=35$, $T=3/2$ quartet shows a discrepancy from the accepted quadratic
form of the IMME with a  coefficient $d=-3.2\,(1.1)$\,keV for the cubic term. On the one hand
questionable experimental data for excited $^{35}$Cl levels have been found in the literature, and
predictions based on isospin-mixing dependent models indicate some possible
deviation, too. Moreover, recent experimental data identified isospin mixing effects in the
vicinity of the $A=35$, $T=3/2$ quartet. From the theoretical calculations and experimental data a
non-zero coefficient $d$ or higher terms are also possible. Further experimental
investigations and a data recheck are needed to confirm this new `breakdown' of the IMME. For example a
direct measurement of the $^{35}$Ar ground state should be performed with the Penning trap
technique. Additional decay studies and spin-assignment checks for the $^{35}$Cl and $^{35}$Ar
mirror nuclides should be performed in order to find new isospin-mixing effects for lower spin
levels. Finally, new challenges are opened to theoretical calculations in order to reproduce the
experimental data with better precision.

\begin{acknowledgments}
The authors would like to thank A.~Brown, A.~Garc\'{\i}a, and P.~Van~Isacker for fruitful discussions.
This work was supported by the German Ministry for Education and Research (BMBF) under contract
06GF151 and 06MZ215, the European Commission within the NIPNET RTD and EURONS/TRAPSPEC networks
under contract HPRI-CT-2001-50034 and RII3-CT-2004-506065, respectively. C.Y. was funded by the
European Commission under the Marie Curie Fellowship network HPMT-CT-2000-00197. K.B. and S.G. are
funded by the Helmholtz association for national research centers (HGF) under contract number
VH-NG-037.
\end{acknowledgments}

%

\bibliographystyle{alpha}


\end{document}